\documentclass[sigconf,anonymous=false]{acmart}

\usepackage{algorithmic}
\usepackage{graphicx}
\usepackage{textcomp}
\usepackage{xcolor}
\usepackage{makecell}

\usepackage{multirow}
\usepackage{balance}

\usepackage{siunitx}
\usepackage{pifont}
\usepackage{bm}

\usepackage[toc]{appendix}

\usepackage[ruled,%
            linesnumbered]{algorithm2e}
\SetKwInput{KwInput}{Input}

\AtBeginDocument{%
  }

\copyrightyear{2025}
\acmYear{2025}
\setcopyright{acmlicensed}
\acmConference[UbiComp/ISWC '25]{Companion of the 2024 ACM International Joint Conference on Pervasive and Ubiquitous Computing}{October 12--16, 2025}{Espoo, Finland} \acmBooktitle{Companion of the 2025 ACM International Joint Conference on Pervasive and Ubiquitous Computing (UbiComp Companion '25s), October 12--16, 2025, Espoo, Finland}
\acmDOI{10.1145/3675094.3678xxx}
\acmISBN{979-8-4007-1058-2/24/xx}

\usepackage{fancyhdr}   

\fancypagestyle{firstpage}{     
  \fancyhf{}
  \chead{\textcolor{gray}{This article has been accepted for publication in the proceedings of the UbiComp/ISWC'25 Adjunct: Adjunct Proceedings of the 2025 ACM International Joint Conference on Pervasive and Ubiquitous Computing \& the 2025 ACM International Symposium on Wearable Computing}}
  \fancyfoot[C]{\small{\textcolor{gray}{~\copyright~ 2025 ACM.  Personal use of this material is permitted.  Permission from ACM must be obtained for all other uses, in any current or future media, including reprinting/republishing this material for advertising or promotional purposes, creating new collective works, for resale or redistribution to servers or lists, or reuse of any copyrighted component of this work in other works.}}}

}

\begin{document}

\title{Passive Body-Area Electrostatic Field (Human Body Capacitance) for Ubiquitous Computing}

\author{Sizhen Bian}
\affiliation{%
  \institution{DFKI}
  \city{Kaiserslautern}
  \country{Germany}
}
\email{sizhen.bian@dfki.de}

\author{Mengxi Liu}
\affiliation{%
  \institution{DFKI}
  \city{Kaiserslautern}
  \country{Germany}
}
\email{mengxi.liu@dfki.de}

\author{Paul Lucowicz}
\affiliation{%
  \institution{DFKI}
  \city{Kaiserslautern}
  \country{Germany}
}
\email{paul.lukowicz@dfki.de}

\renewcommand{\shortauthors}{Sizhen Bian et al.}

\begin{abstract}

Passive body-area electrostatic field sensing, also referred to as human body capacitance (HBC), is an energy-efficient and non-intrusive sensing modality that exploits the human body’s inherent electrostatic properties to perceive human behaviors. 
This paper presents a focused overview of passive HBC sensing, including its underlying principles, historical evolution, hardware architectures, and applications across research domains. Key challenges, such as susceptibility to environmental variation, are discussed to trigger mitigation techniques. Future research opportunities in sensor fusion and hardware enhancement are highlighted. To support continued innovation, this work provides open-source resources and aims to empower researchers and developers to leverage passive electrostatic sensing for next-generation wearable and ambient intelligence systems.

\end{abstract}



\begin{teaserfigure}
  \includegraphics[width=0.99\linewidth, height = 11.5cm]{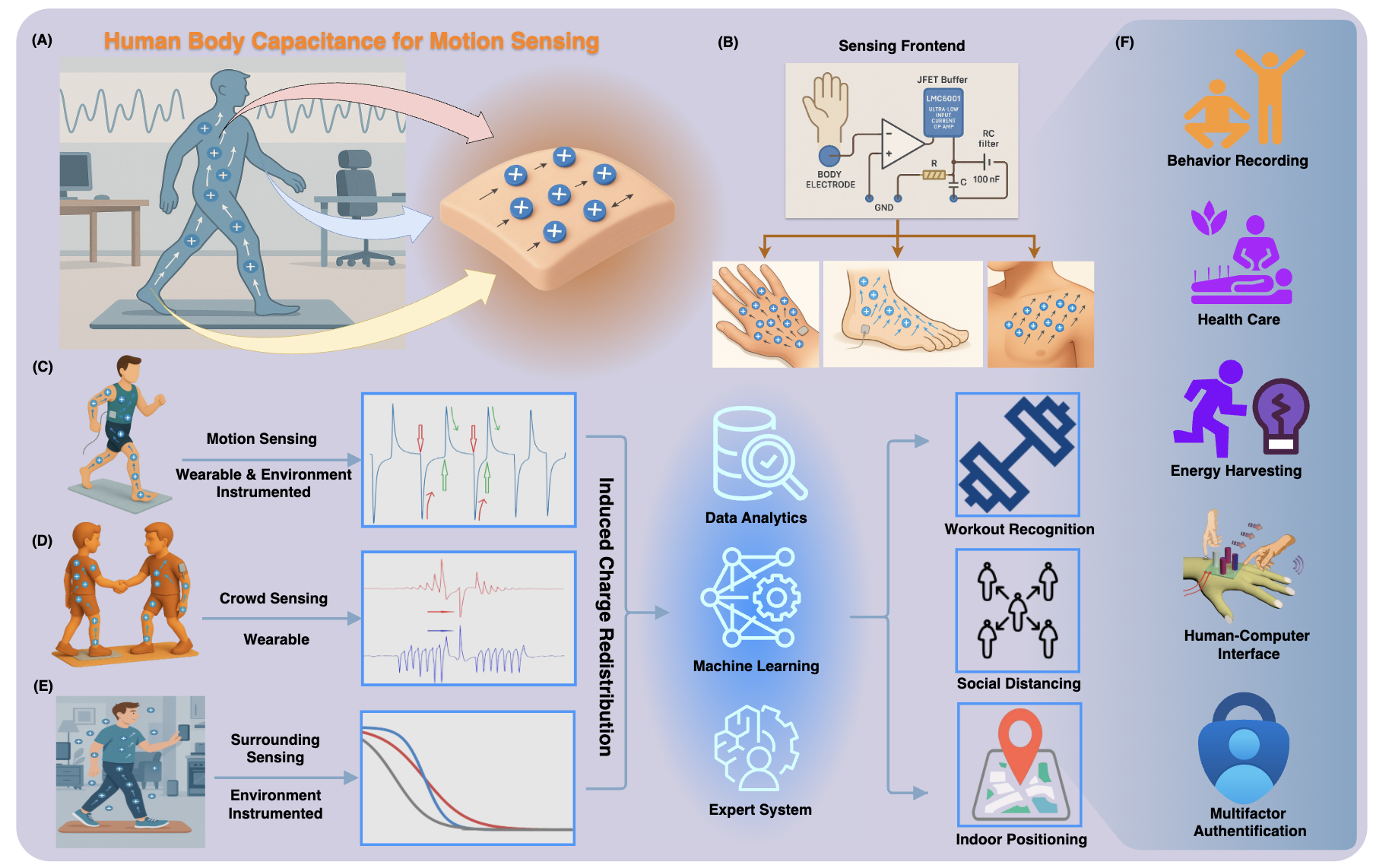}
    \caption{Human Body Capacitance-based motion sensing for human activity recognition. (A) Principle: redistribution of body surface charge due to capacitance variation during movement. (B) Amplifier-based sensing frontend for capturing surface charge dynamics. (C–E) Example applications of HBC-based motion/crowd/environment sensing using wearable or ambient-instrumented sensing units, along with corresponding signal processing pipelines, enabling tasks such as workout recognition, social distancing, and indoor localization. (F) Overview of representative studies across multiple research domains.}
    \label{Overview}
\end{teaserfigure}


\maketitle

\section{Introduction}

\thispagestyle{firstpage}

The ability to sense and interpret human body motion is fundamental to a wide range of applications, including human activity recognition (HAR), human–computer interaction (HCI), healthcare monitoring, and smart environments. Traditional sensing modalities—such as inertial measurement units (IMUs), cameras, and physiological sensors—have enabled significant advances in these areas. However, each comes with limitations: inertial sensors often require direct attachment and are prone to drift \cite{zhao2018review}, vision-based systems raise privacy concerns and depend on line-of-sight \cite{cartas2020activities}, and physiological sensors can be obtrusive or power-intensive \cite{casson2019wearable}. These challenges have driven interest in alternative sensing paradigms that are unobtrusive, low-power, and privacy-preserving.

One promising and underexplored modality is passive body-area electrostatic field sensing, also known as human body capacitance (HBC \cite{jonassen1998human}) sensing. This approach leverages the natural electrostatic field generated by the human body and its interactions with nearby objects and environments. Unlike active capacitance sensing systems that require external signal excitation \cite{bian2021capacitive, liu2022non,zhou2023mocapose，bian2021systematic}, passive HBC sensing operates without the need for additional signal sources. Instead, it detects subtle charge redistributions on the body surface that occur during movement, enabling a highly power-efficient and non-intrusive method for capturing motion-related information.
Passive HBC sensing exhibits several unique properties. Its sensitivity to charge variation allows it to detect body motion across different body parts, including non-instrumented limbs, enabling cross-body part sensing from a single sensor location \cite{bian2022using}. It also facilitates interaction detection in ambient environments, making it well suited for applications involving proximity, posture, gesture, or group activity \cite{suh2023worker,bian2024body}. These characteristics position passive HBC sensing as a compelling alternative or complement to conventional motion sensing technologies.

Despite its potential, passive HBC sensing remains relatively underdeveloped in terms of systematization and widespread deployment. The sensing signals are often weak and susceptible to environmental noise, and the physical mechanisms underlying surface charge redistribution are not yet fully characterized. Moreover, the field lacks a unified understanding of system design trade-offs, hardware implementations, and best practices for signal processing and application development. 
This work is thus drafted towards addressing these gaps by providing a focused overview of passive HBC sensing for body motion detection. We begin by introducing the physical principles and historical background of electrostatic body-area sensing. We then describe key frontend architectures that enable reliable surface charge measurement. Practical applications across domains are then briefly reviewed, followed by a discussion of system limitations, including signal conditioning and environmental dependency. Finally, we outline future directions for research and development, emphasizing opportunities in multi-modal sensing, hardware optimization, and open-source ecosystem building. By consolidating knowledge across disciplines, this paper serves as both a primer and a roadmap for researchers and practitioners interested in passive electrostatic sensing and accelerates the exploration and adoption of human body capacitance in ubiquitous computing systems.


\section{Sensing Principle}

\begin{figure}[hbt]
\centering
\includegraphics[width=0.79\linewidth, height = 4.0cm]{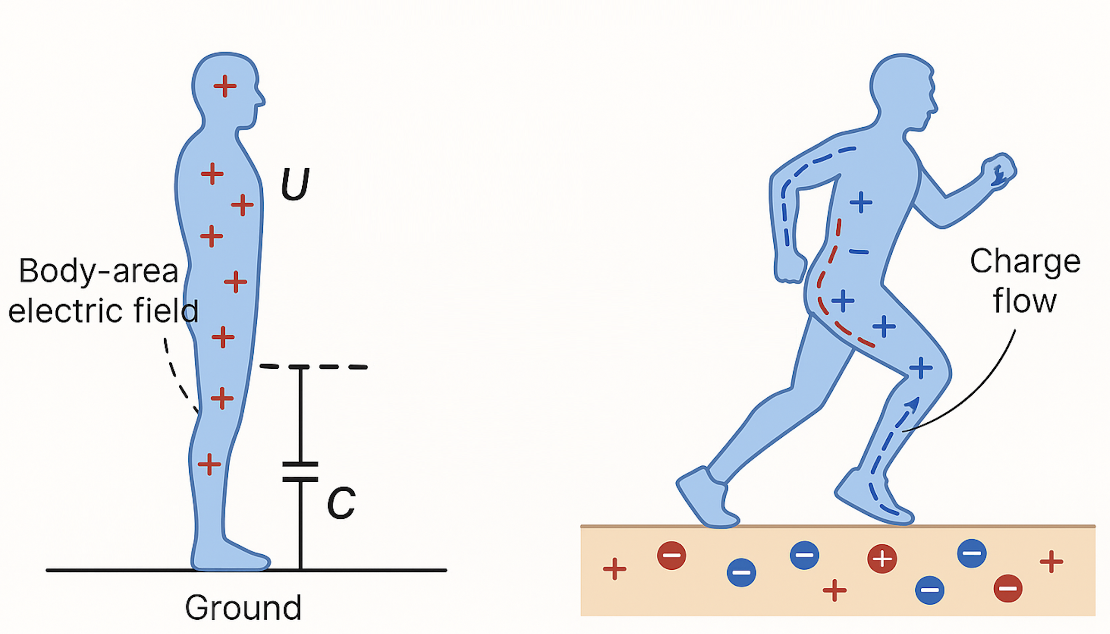}
\caption{Sensing Principle}
\label{Principle}
\end{figure}

The passive body-area electrostatic field, also referred to as human body capacitance (HBC), arises from the static electric field naturally formed between the human body and its surrounding environment—particularly the ground \cite{jonassen1998human}. This field results from charge imbalances that develop as the body, which is predominantly water-based (~60\%) and therefore conductive, becomes insulated from the environment (e.g., by clothing or footwear). This insulation allows the body to maintain a surface potential relative to its surroundings, forming a passive, quasi-static electric field in the immediate body area.
The human body in this context can be modeled as a floating conductor in a capacitor-like system, as shown in Fig. \ref{Principle} (left). The surface potential U, total surface charge Q, and capacitance 
C together define the electrical state of the body. The ground (or nearby objects) acts as the other side of the capacitor. When the body moves, such as during walking or gesturing, changes in relative distance, orientation, or overlapping area between the body and surroundings cause dynamic variations in the body’s capacitance C. This, in turn, induces a redistribution of surface charge Q, manifesting as weak charge flows on the skin’s surface. These temporal variations in surface potential and induced charge are collectively described as the body-area electrostatic field dynamics.
By capturing these minute charge flow patterns, it becomes possible to infer body motion without active signal injection. This passive sensing method offers multiple advantages: it is extremely low-power, non-invasive, and cost-effective, and it allows for position-free deployment, meaning sensors can often detect full-body motion even when placed at a single point (e.g., wrist or ankle).

\section{Historical Background}

The concept of capacitance has been intertwined with the history of electricity since its earliest exploration. One of the first capacitive devices, the Leyden jar, was invented in the 18th century to store static electric charge, during a time when electricity was still imagined as a fluid-like phenomenon \cite{heilbron1966gm}. In the 1830s, Michael Faraday conducted foundational experiments demonstrating that the material between a capacitor’s plates influenced its ability to store charge — a discovery that led to the naming of the Farad as the unit of capacitance. Over the ensuing decades, diverse capacitor types were developed to support industrial demands, including ceramic capacitors for resonant circuits and tantalum capacitors for transistor support in low-voltage applications.
A key historical milestone in body capacitance sensing came with the invention of the Theremin in 1919 by Leon Theremin, a 23-year-old Russian physicist and musician. Originally developed by accident, the Theremin became the first consumer device to exploit body capacitance to control a musical tone via hand proximity \cite{nikitin2012leon}. This instrument marked the birth of human body capacitance (HBC)-enabled interaction and has since been widely cited in electrostatic sensing literature \cite{smith1996field, grosse2017finding, braun2015capacitive}. Like the early aviation pioneers, Theremin’s work was driven by experimentation rather than established scientific models—an analogy that captures the exploratory spirit of early HBC research.
With time, the electrical theory behind the Theremin was formalized, and digital implementations emerged \cite{kuik2004digital}, supported by verified physical models of human body electrostatics \cite{skeldon1998physics}. Interestingly, nature itself offers parallels: electrosensitive fish such as Gymnarchus niloticus exhibit remarkable capabilities akin to HBC. These species detect variations in the surrounding electric field via distributed surface receptors, enabling them to identify nearby objects with differing electrical properties \cite{lissmann1958mechanism, bullock1982electroreception}.
In the biomedical domain, body-area capacitance measurements became relevant in clinical settings. Work by Jonassen et al. \cite{jonassen1998human} and Forster et al. \cite{forster1974measurement} investigated human body capacitance in hospital environments, particularly in isolated patients. These studies highlighted how ambient electrostatic fields from unshielded wiring could affect patient safety due to unintended capacitive coupling with medical equipment.
Entering the 21st century, HBC sensing experienced a significant surge in development driven by advances in microelectronics, low-power chip design, and ubiquitous computing. Modern resonator-based capacitance sensors (e.g., Texas Instruments’ FDC2x1x series) and charge-variation-based sensors (e.g., STMicroelectronics’ QVAR) have enabled accurate, high-resolution HBC sensing in compact, wearable formats. These developments have unlocked a wide range of applications, including: Biophysical signal monitoring \cite{yama2007development}, Proximity and position tracking \cite{osoinach2007proximity}, Human activity classification\cite{cheng2010active, bian2019passive}, Intra-body communication systems \cite{shinagawa2004near}, as summarized in the following section.

\section{Applications}

\begin{table*}[htbp]
\caption{Passive body-area electric field studies}
\label{BodyPartTable}
\footnotesize
\begin{tabular}{ p{1.0cm} p{1.0cm} p{2.7cm} p{1.5cm} p{2.2cm}  p{6.8cm}}
\toprule
Work  & Prototype & Subject & Source Signal & Hardware & Performance \\ 
\midrule
\cite{takiguchi2007human}-2007 & Environment beacon &  Step Counting & Current &  Amplification circuit  &   Detecting the number of steps at an accuracy of approximately 99.4\%.
\\ 
\cite{cheng2008body}-2008 & Clothes & Control gesture Recognition & Frequency  & MC34940 & Three hand control gestures with an accuracy of around 89\% 
\\
\cite{braun2009using}-2009  & Off-body patch  & Gesture Detection, etc  & Current  &  CY3235 & Qualitative analysis 
\\ 
\cite{ianov2012development}-2012   & Clothes & Non-contact bioelectrical signal detection  & Current  & Instrumen-tation Amplifier & Capable of correctly recording ECG and EMG under different loads 
\\
\cite{grosse2012enhancing}-2012  & Wrist-band          & Activity Recognition &  Frequency  & 555 Timer & Nine everyday-life activities with an F-score of 73.5\% 
\\
\cite{cohn2012ultra}-2012 & Wrist- band & Body motion sensing &  Current & General amplifier & Nearly 92\% classification accuracy of rest, small arm movements, walking, and jogging. 
\\
\cite{braun2013capacitive}-2013 & Off-body  & Hand festure recognition  & Current  &  CY3271 & By only a handful gestures, the system shows reliable basic home control 
\\ 
\cite{braun2014towards}-2014    & Armrest         & Hand Gesture Recognition    & Current  &  TMS320F2 & Detection rate between 77.3\% and 90.9\% for touch set and free-air set between 45.5\%  and 81.8\% 
\\ 
\cite{choi2017driver}-2017      & Seat         & Non-contact ECG Detection  & Current  &  cECG & An averaged correlation coefficient value between the ture movement signal and ECG-decomposed movement signals was 0.77 
\\ 
\cite{matthies2017earfieldsensing}-2017        & Earbud        & Facial Expression Recognition    & Current  & Discrete component, Amplifier & 5 facial gestures with a precision of 90\% while sitting and 85.2\% while walking  
\\ 
\cite{erickson2018tracking}-2018   & Robot Arm  & Robot assistant living & Current  &  MPR121 & The robot successfully
pulled the sleeve of a hospital gown and a cardigan onto the right
arms of 10 human participants. 
\\ 
\cite{tang2019indoor}-2019 & Environment beacon & Indoor occupancy Awareness/ Localization  & current & EPS & human identification with an accuracy of 98.3\%, occupant localization accuracy of 89.03\% with average error less than 0.3 m. 
\\
\cite{bian2019wrist}-2019 & Wristband & Collaborative Activity Detection &  Current & Customized front end  & One-shot collaborative signal analysis.
\\ 
\cite{bian2022using}-2022 & Wrist- band  & Leg-exercise recognition & Current & Customized front end & Body capacitance performs better than IMU in classification(0.89 vs 0.78 F-score) and counting. 
\\ 
\cite{bian2022contribution}-2022 & Wristband &Collaborative Activity Detection  & Current & Customized front end  &  The capacitive sensor can
improve the recognition of collaborative activities with an F-score over a single wrist accelerometer
approach by 16\%. 
\\ 
\cite{bian2022human}-2022 & Wristband & Social Distancing & Current & Customized front end  & A true positive alert rate of 74.3\% when tests was performed in an indoor environment and only 46\% in an outdoor environment.
\\ 
\cite{bian2024earable}-2024 & Earbud/ Wristband &  Step Counting & Current &  Qvar &   Better accuracy than the commercial wrist-worn devices (e.g.,96\% vs. 66\% of the Fitbit when walking in the shopping center while pushing a shopping trolley).
\\ 
\bottomrule
\end{tabular}
\end{table*}

Human Body Capacitance (HBC) sensing has demonstrated its versatility across a wide array of application domains. To provide a comprehensive view of recent advancements, Table \ref{BodyPartTable} summarizes representative HBC studies, highlighting the sensing configurations, subjects, signal sources, and target applications.
\paragraph{\textbf{Basic Motion Sensing and Step Counting.}}
Early applications of HBC sensing focused on detecting gross body movements such as walking or stepping. For instance, one of the earliest implementations \cite{takiguchi2007human} achieved 99.4\% accuracy in step counting using an environmental beacon and an amplification circuit. More recent work \cite{bian2024earable} employed a commercial Qvar chip on an earbud or wristband to achieve higher accuracy than commercial fitness trackers, particularly in particular contexts such as pushing a shopping trolley. These studies underscore HBC’s advantage in motion tasks where traditional IMUs suffer from placement sensitivity.

\paragraph{\textbf{Gesture and Expression Recognition.}}
HBC sensors are highly responsive to subtle changes in body posture and proximity, making them suitable for gesture recognition and facial expression detection. Studies such as \cite{cheng2008body} and \cite{braun2013capacitive} showed reliable performance in hand gesture control, achieving accuracies up to 89\%. A notable application \cite{matthies2017earfieldsensing} demonstrated facial gesture recognition using earbuds, with over 90\% precision while seated and 85.2\% while walking, enabling discreet emotion-sensing capabilities for wearable HCI.

\paragraph{\textbf{Human–Computer Interaction (HCI) and Assistive Technologies.}}
Several works explored the integration of HBC sensors into HCI systems and assistive robotics. For example, \cite{erickson2018tracking} used a robotic arm to assist with dressing tasks, identifying sleeve and gown locations via capacitance variations on human limbs. Likewise, \cite{braun2014towards} and \cite{ianov2012development} explored gesture and bioelectrical signal detection for interaction with digital systems. These applications benefit from HBC’s non-contact, low-latency sensing capabilities and capacity to detect movement in clothed or partially obstructed conditions.

\paragraph{\textbf{Activity Recognition and Behavioral Analysis.}}
HBC has also been employed in human activity recognition (HAR), especially for classifying movement types and understanding context. Wristband-based systems \cite{cohn2012ultra}, and \cite{bian2022using} reported classification accuracies over 89\% for various daily movements, with \cite{bian2022contribution} showing a 16\% improvement in collaborative activity recognition compared to accelerometer-only approaches. This highlights HBC’s ability to capture whole-body movement patterns through localized sensors.

\paragraph{\textbf{Social and Environmental Awareness.}}
Beyond individual motion tracking, HBC sensing has been extended to environmental and social context detection. For example, \cite{tang2019indoor} employed environmental beacons to enable indoor localization and occupancy detection, achieving human identification accuracies of up to 98.3\% with less than 0.3 m localization error. More recently, \cite{bian2022human} used wrist-worn HBC sensors to detect social distancing violations, achieving a true positive rate of 74.3\% indoors, demonstrating the potential of HBC in public health and safety monitoring.

The wide range of applications—from step counting, gesture recognition, and collaborative behavior sensing to environmental awareness and assistive robotics—demonstrates that passive HBC sensing is a flexible and high-impact modality. Its core advantages—contactless operation, position independence, and low power requirements—enable seamless integration into daily-life wearable and ambient systems. 
Ongoing advancements in frontend design, sensor integration, and signal processing continue to expand the potential of HBC-based technologies.

\section{Hardware Implementation}

\begin{figure}
\begin{minipage}[t]{0.45\linewidth}
\centering
\includegraphics[width=0.89\textwidth,height=3.0cm]{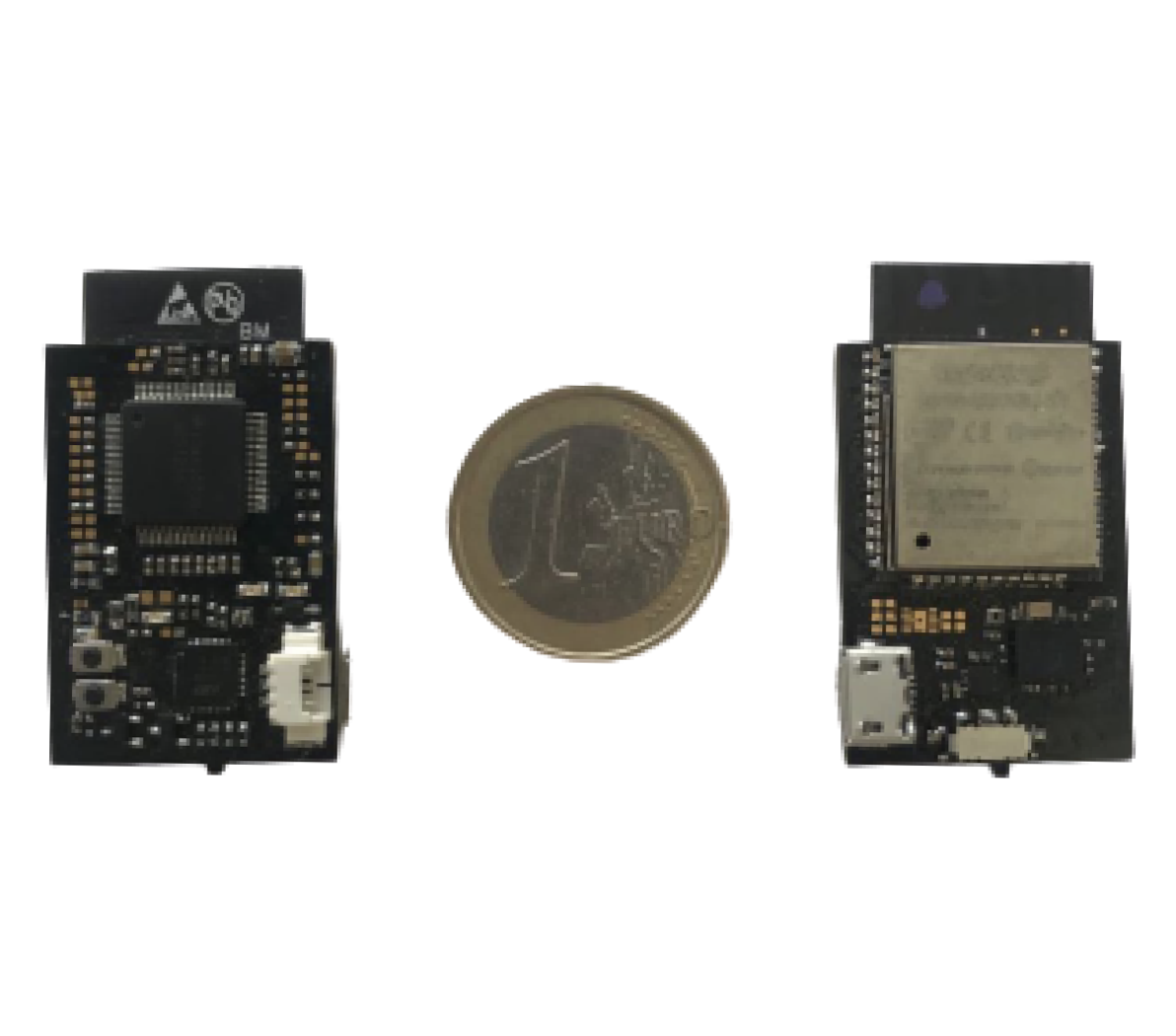}
\caption{Sensing frontend with 24-bit ADC and two divider resistors \cite{HighResolutionADC_for_HBC}}
\label{Sensor}
\end{minipage}
\quad
\begin{minipage}[t]{0.45\linewidth}
\centering
\includegraphics[width=0.89\textwidth,height=3.0cm]{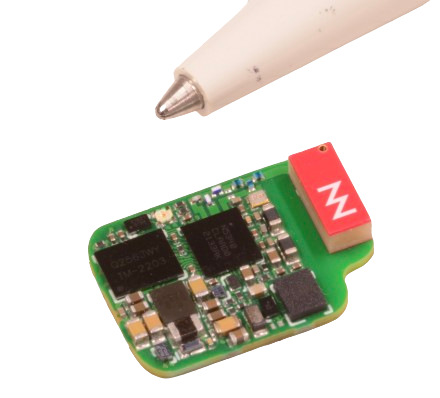}
\caption{Sensing frontend with integrated Qvar module \cite{VitalPod}}
\label{Prototype}
\end{minipage}
\end{figure}

\begin{figure*}[hbt]
\centering
\includegraphics[width=0.95\linewidth, height = 20.5cm]{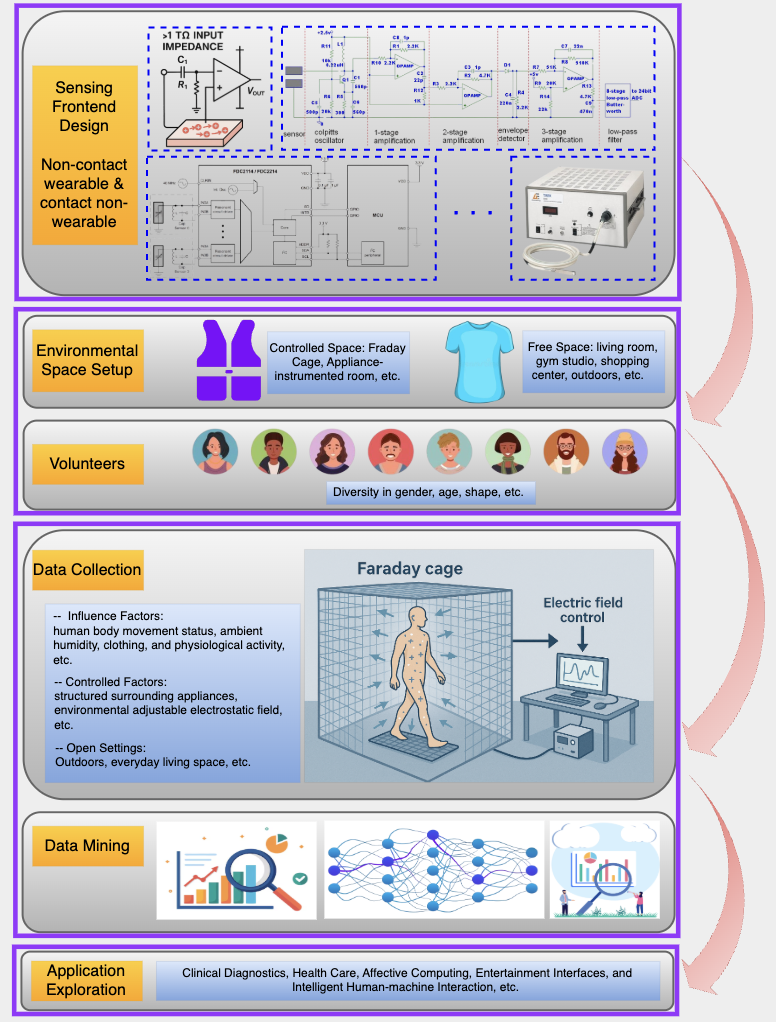}
\caption{Working Guidelines of HBC Studies in Ubiquitous Computing}
\label{Guidline}
\end{figure*}

The effectiveness of passive human body capacitance (HBC) sensing relies heavily on the careful design of the sensing hardware. Because the electrostatic signals involved are extremely weak and sensitive to environmental influences, hardware implementations must strike a balance between signal fidelity, power efficiency, and deployment flexibility. This section outlines the typical components and configurations used in HBC sensing systems:

\paragraph{\textbf{Electrodes.}}
At the core of any HBC sensing system are the electrodes, which serve as the primary interface for capturing the body’s surface potential or capacitive coupling with the environment. Electrodes can be embedded in wearable form factors (e.g., wristbands, clothing) or integrated into ambient infrastructure (e.g., seats, walls, or objects). Depending on the application, electrodes may be placed in direct contact with the skin or used in a non-contact, coupled configuration, leveraging capacitive coupling through insulating layers such as clothing.

\paragraph{\textbf{Signal Acquisition.}}
The weak nature of passive electrostatic signals necessitates careful signal acquisition strategies. Three primary approaches are commonly used: 
\newline\textbf{A, High-Impedance Amplification Front Ends}: These circuits typically include ultra-high input impedance op-amps (e.g., LMC6001) to prevent signal loading and preserve the integrity of the sensed signal. This configuration is widely adopted for current-mode HBC sensing, where subtle charge redistribution patterns are captured from the body surface. An integrated solution is also available like Qvar from ST Microelectronics (a PCB design is open-sourced as depicted in Fig. \ref{Prototype}). \newline\textbf{B, High-Resolution Analog-to-Digital Converters (ADCs)}:
In systems aiming to directly capture surface voltage or potential differences, high-resolution ADCs (e.g., ADS1298, commonly used in biomedical signal acquisition) are used to digitize signals with high sensitivity and ultra-low noise (an example open-sourced design is depicted in Fig. \ref{Sensor}). \newline\textbf{C, Timing-Based Circuits}: Circuits such as RC timers or 555 timer-based oscillators can be employed to infer capacitance changes indirectly by measuring the charge/discharge time or frequency shift. These are commonly found in frequency-domain HBC sensing systems.

\paragraph{\textbf{Processing unit.}}
The processing unit in an HBC system is typically a lightweight embedded microcontroller (e.g., ARM Cortex-M, ATmega series), responsible for signal processing, data logging, and wireless transmission. These units may locally handle real-time signal filtering, feature extraction, and simple decision logic or transmit data to a more capable device for advanced analysis (e.g., smartphone or cloud server).\newline

Among the above frontends, two major signal domains are used and each with unique trade-offs:

\noindent\textbf{A, Current-Based Mode:}
This is the most commonly used mode in low-power applications. It captures charge or voltage variations directly from the body. While power-efficient and relatively simple to implement, this mode is more sensitive to environmental noise and variations in grounding or insulation.

\noindent\textbf{B, Frequency-Based Mode:}
In this approach, capacitance changes are translated into frequency shifts using oscillators or timer-based circuits. Though potentially more robust to environmental variation, frequency-based systems often require more complex circuitry and calibration to maintain accuracy.

\section{Limitations and Future Work}

While HBC sensing offers compelling advantages such as deployment flexibility, it faces certain limitations. These challenges must be addressed to enable robust, scalable adoption of HBC systems in real-world applications.

\paragraph{\textbf{Environmental Sensitivity:}}
Passive capacitive sensing is inherently susceptible to variations in the surrounding electrostatic environment. Fluctuations in ambient electric fields, temperature, or humidity can distort signal baselines, introducing drift or inconsistency in measurements \cite{}.

\paragraph{\textbf{Context-Dependent Performance:}}
External factors such as footwear, clothing material, and environmental surfaces can alter the body’s capacitive properties. These variations often result in classification degradation or require model retraining for different deployment settings \cite{bian2021systematic}.

\paragraph{\textbf{Limited Generalization Across Users and Conditions:}}
Models trained on specific individuals or environmental configurations may fail to generalize to new users, poses, or contexts. This is particularly problematic for large-scale or personalized applications without recalibration \cite{kang2024device}.\newline

To overcome the above limitations, several promising directions are emerging:

\paragraph{\textbf{Sensor Fusion:}}
Combining HBC with other sensing modalities (e.g., inertial measurement units, optical sensors, or bioelectrical signals) offers a pathway to enhance resolution, contextual awareness, and system resilience.

\paragraph{\textbf{Hardware Enhancements:}}
Advances in active shielding, differential signal acquisition, and application-specific electrode design can improve signal quality and reduce environmental interference.

\paragraph{\textbf{Standardization and Benchmarking:}}
Developing common datasets, evaluation metrics, and open-source platforms will help unify research efforts and accelerate progress in the field.

\section{Conclusion}

As HBC sensing continues to mature, its inherent advantages — contactless operation, low power consumption, cost efficiency, and deployment flexibility — position it as a compelling alternative to traditional sensing modalities in HAR and HCI (working guidelines of HBC-based studies are depicted as Fig. \ref{Guidline}). This paper has reviewed the fundamental principles, hardware implementations, application domains, and ongoing challenges of passive HBC systems. As the field continues to evolve, addressing limitations will be critical. With continued innovation in sensing hardware, signal processing, and multi-modal integration, HBC is well-positioned to play an essential role in next-generation wearable and ambient intelligence systems.




\bibliographystyle{ACM-Reference-Format}
\balance
\bibliography{sample-base}


\begin{thebibliography}{44}


\ifx \showCODEN    \undefined \def \showCODEN     #1{\unskip}     \fi
\ifx \showDOI      \undefined \def \showDOI       #1{#1}\fi
\ifx \showISBNx    \undefined \def \showISBNx     #1{\unskip}     \fi
\ifx \showISBNxiii \undefined \def \showISBNxiii  #1{\unskip}     \fi
\ifx \showISSN     \undefined \def \showISSN      #1{\unskip}     \fi
\ifx \showLCCN     \undefined \def \showLCCN      #1{\unskip}     \fi
\ifx \shownote     \undefined \def \shownote      #1{#1}          \fi
\ifx \showarticletitle \undefined \def \showarticletitle #1{#1}   \fi
\ifx \showURL      \undefined \def \showURL       {\relax}        \fi
\providecommand\bibfield[2]{#2}
\providecommand\bibinfo[2]{#2}
\providecommand\natexlab[1]{#1}
\providecommand\showeprint[2][]{arXiv:#2}

\bibitem[Hig(2022)]%
        {HighResolutionADC_for_HBC}
 \bibinfo{year}{2022}\natexlab{}.
\newblock \bibinfo{title}{High Resolution ADC with two divider resistors for HBC}.
\newblock \bibinfo{howpublished}{\url{https://github.com/zhaxidele/Toolkit-for-HBC-sensing/tree/main}}.
\newblock
\newblock
\shownote{Accessed: 2025-04-30}.


\bibitem[Bian(2022)]%
        {bian2022human}
\bibfield{author}{\bibinfo{person}{Sizhen Bian}.} \bibinfo{year}{2022}\natexlab{}.
\newblock \emph{\bibinfo{title}{Human Activity Recognition with Field Sensing Technique}}.
\newblock \bibinfo{thesistype}{Ph.\,D. Dissertation}. \bibinfo{school}{Technische Universit{\"a}t Kaiserslautern}.
\newblock


\bibitem[Bian et~al\mbox{.}(2024a)]%
        {bian2024body}
\bibfield{author}{\bibinfo{person}{Sizhen Bian}, \bibinfo{person}{Mengxi Liu}, \bibinfo{person}{Bo Zhou}, \bibinfo{person}{Paul Lukowicz}, {and} \bibinfo{person}{Michele Magno}.} \bibinfo{year}{2024}\natexlab{a}.
\newblock \showarticletitle{Body-Area Capacitive or Electric Field Sensing for Human Activity Recognition and Human-Computer Interaction: A Comprehensive Survey}.
\newblock \bibinfo{journal}{\emph{Proceedings of the ACM on Interactive, Mobile, Wearable and Ubiquitous Technologies}} \bibinfo{volume}{8}, \bibinfo{number}{1} (\bibinfo{year}{2024}), \bibinfo{pages}{1--49}.
\newblock


\bibitem[Bian and Lukowicz(2021a)]%
        {bian2021capacitive}
\bibfield{author}{\bibinfo{person}{Sizhen Bian} {and} \bibinfo{person}{Paul Lukowicz}.} \bibinfo{year}{2021}\natexlab{a}.
\newblock \showarticletitle{Capacitive sensing based on-board hand gesture recognition with TinyML}. In \bibinfo{booktitle}{\emph{Adjunct Proceedings of the 2021 ACM International Joint Conference on Pervasive and Ubiquitous Computing and Proceedings of the 2021 ACM International Symposium on Wearable Computers}}. \bibinfo{pages}{4--5}.
\newblock


\bibitem[Bian and Lukowicz(2021b)]%
        {bian2021systematic}
\bibfield{author}{\bibinfo{person}{Sizhen Bian} {and} \bibinfo{person}{Paul Lukowicz}.} \bibinfo{year}{2021}\natexlab{b}.
\newblock \showarticletitle{A systematic study of the influence of various user specific and environmental factors on wearable human body capacitance sensing}. In \bibinfo{booktitle}{\emph{EAI International Conference on Body Area Networks}}. Springer, \bibinfo{pages}{247--274}.
\newblock


\bibitem[Bian et~al\mbox{.}(2019a)]%
        {bian2019passive}
\bibfield{author}{\bibinfo{person}{Sizhen Bian}, \bibinfo{person}{Vitor~F Rey}, \bibinfo{person}{Peter Hevesi}, {and} \bibinfo{person}{Paul Lukowicz}.} \bibinfo{year}{2019}\natexlab{a}.
\newblock \showarticletitle{Passive capacitive based approach for full body gym workout recognition and counting}. In \bibinfo{booktitle}{\emph{2019 IEEE International Conference on Pervasive Computing and Communications (PerCom}}. IEEE, \bibinfo{pages}{1--10}.
\newblock


\bibitem[Bian et~al\mbox{.}(2019b)]%
        {bian2019wrist}
\bibfield{author}{\bibinfo{person}{Sizhen Bian}, \bibinfo{person}{Vitor~F Rey}, \bibinfo{person}{Junaid Younas}, {and} \bibinfo{person}{Paul Lukowicz}.} \bibinfo{year}{2019}\natexlab{b}.
\newblock \showarticletitle{Wrist-worn capacitive sensor for activity and physical collaboration recognition}. In \bibinfo{booktitle}{\emph{2019 IEEE International Conference on Pervasive Computing and Communications Workshops (PerCom Workshops)}}. IEEE, \bibinfo{pages}{261--266}.
\newblock


\bibitem[Bian et~al\mbox{.}(2022a)]%
        {bian2022contribution}
\bibfield{author}{\bibinfo{person}{Sizhen Bian}, \bibinfo{person}{Vitor~Fortes Rey}, \bibinfo{person}{Siyu Yuan}, {and} \bibinfo{person}{Paul Lukowicz}.} \bibinfo{year}{2022}\natexlab{a}.
\newblock \showarticletitle{The Contribution of Human Body Capacitance/Body-Area Electric Field To Individual and Collaborative Activity Recognition}.
\newblock \bibinfo{journal}{\emph{arXiv preprint arXiv:2210.14794}} (\bibinfo{year}{2022}).
\newblock


\bibitem[Bian et~al\mbox{.}(2024b)]%
        {bian2024earable}
\bibfield{author}{\bibinfo{person}{Sizhen Bian}, \bibinfo{person}{Rakita Strahinja}, \bibinfo{person}{Philipp Schilk}, \bibinfo{person}{Cl{\'e}nin Marc-Andr{\'e}}, \bibinfo{person}{Silvano Cortesi}, \bibinfo{person}{Kanika Dheman}, \bibinfo{person}{Elio Reinschmidt}, {and} \bibinfo{person}{Michele Magno}.} \bibinfo{year}{2024}\natexlab{b}.
\newblock \showarticletitle{Earable and Wrist-worn Setup for Accurate Step Counting Utilizing Body-Area Electrostatic Sensing}. In \bibinfo{booktitle}{\emph{Companion of the 2024 on ACM International Joint Conference on Pervasive and Ubiquitous Computing}}. \bibinfo{pages}{904--910}.
\newblock


\bibitem[Bian et~al\mbox{.}(2022b)]%
        {bian2022using}
\bibfield{author}{\bibinfo{person}{Sizhen Bian}, \bibinfo{person}{Siyu Yuan}, \bibinfo{person}{Vitor~Fortes Rey}, {and} \bibinfo{person}{Paul Lukowicz}.} \bibinfo{year}{2022}\natexlab{b}.
\newblock \showarticletitle{Using human body capacitance sensing to monitor leg motion dominated activities with a wrist worn device}.
\newblock In \bibinfo{booktitle}{\emph{Sensor-and Video-Based Activity and Behavior Computing}}. \bibinfo{publisher}{Springer}, \bibinfo{pages}{81--94}.
\newblock


\bibitem[Braun et~al\mbox{.}(2013)]%
        {braun2013capacitive}
\bibfield{author}{\bibinfo{person}{Andreas Braun}, \bibinfo{person}{Tim Dutz}, {and} \bibinfo{person}{Felix Kamieth}.} \bibinfo{year}{2013}\natexlab{}.
\newblock \showarticletitle{Capacitive sensor-based hand gesture recognition in ambient intelligence scenarios}. In \bibinfo{booktitle}{\emph{Proceedings of the 6th International Conference on PErvasive Technologies Related to Assistive Environments}}. \bibinfo{pages}{1--4}.
\newblock


\bibitem[Braun and Hamisu(2009)]%
        {braun2009using}
\bibfield{author}{\bibinfo{person}{Andreas Braun} {and} \bibinfo{person}{Pascal Hamisu}.} \bibinfo{year}{2009}\natexlab{}.
\newblock \showarticletitle{Using the human body field as a medium for natural interaction}. In \bibinfo{booktitle}{\emph{Proceedings of the 2nd International Conference on PErvasive Technologies Related to Assistive Environments}}. \bibinfo{pages}{1--7}.
\newblock


\bibitem[Braun et~al\mbox{.}(2014)]%
        {braun2014towards}
\bibfield{author}{\bibinfo{person}{Andreas Braun}, \bibinfo{person}{Stephan Neumann}, \bibinfo{person}{S{\"o}nke Schmidt}, \bibinfo{person}{Reiner Wichert}, {and} \bibinfo{person}{Arjan Kuijper}.} \bibinfo{year}{2014}\natexlab{}.
\newblock \showarticletitle{Towards interactive car interiors: the active armrest}. In \bibinfo{booktitle}{\emph{Proceedings of the 8th Nordic Conference on Human-Computer Interaction: Fun, Fast, Foundational}}. \bibinfo{pages}{911--914}.
\newblock


\bibitem[Braun et~al\mbox{.}(2015)]%
        {braun2015capacitive}
\bibfield{author}{\bibinfo{person}{Andreas Braun}, \bibinfo{person}{Reiner Wichert}, \bibinfo{person}{Arjan Kuijper}, {and} \bibinfo{person}{Dieter~W Fellner}.} \bibinfo{year}{2015}\natexlab{}.
\newblock \showarticletitle{Capacitive proximity sensing in smart environments}.
\newblock \bibinfo{journal}{\emph{Journal of Ambient Intelligence and Smart Environments}} \bibinfo{volume}{7}, \bibinfo{number}{4} (\bibinfo{year}{2015}), \bibinfo{pages}{483--510}.
\newblock


\bibitem[Bullock(1982)]%
        {bullock1982electroreception}
\bibfield{author}{\bibinfo{person}{Theodore~H Bullock}.} \bibinfo{year}{1982}\natexlab{}.
\newblock \showarticletitle{Electroreception.}
\newblock \bibinfo{journal}{\emph{Annual review of neuroscience}} (\bibinfo{year}{1982}).
\newblock


\bibitem[Cartas et~al\mbox{.}(2020)]%
        {cartas2020activities}
\bibfield{author}{\bibinfo{person}{Alejandro Cartas}, \bibinfo{person}{Petia Radeva}, {and} \bibinfo{person}{Mariella Dimiccoli}.} \bibinfo{year}{2020}\natexlab{}.
\newblock \showarticletitle{Activities of daily living monitoring via a wearable camera: Toward real-world applications}.
\newblock \bibinfo{journal}{\emph{IEEE access}}  \bibinfo{volume}{8} (\bibinfo{year}{2020}), \bibinfo{pages}{77344--77363}.
\newblock


\bibitem[Casson(2019)]%
        {casson2019wearable}
\bibfield{author}{\bibinfo{person}{Alexander~J Casson}.} \bibinfo{year}{2019}\natexlab{}.
\newblock \showarticletitle{Wearable EEG and beyond}.
\newblock \bibinfo{journal}{\emph{Biomedical engineering letters}} \bibinfo{volume}{9}, \bibinfo{number}{1} (\bibinfo{year}{2019}), \bibinfo{pages}{53--71}.
\newblock


\bibitem[Cheng et~al\mbox{.}(2010)]%
        {cheng2010active}
\bibfield{author}{\bibinfo{person}{Jingyuan Cheng}, \bibinfo{person}{Oliver Amft}, {and} \bibinfo{person}{Paul Lukowicz}.} \bibinfo{year}{2010}\natexlab{}.
\newblock \showarticletitle{Active capacitive sensing: Exploring a new wearable sensing modality for activity recognition}. In \bibinfo{booktitle}{\emph{International conference on pervasive computing}}. Springer, \bibinfo{pages}{319--336}.
\newblock


\bibitem[Cheng et~al\mbox{.}(2008)]%
        {cheng2008body}
\bibfield{author}{\bibinfo{person}{Jingyuan Cheng}, \bibinfo{person}{David Bannach}, {and} \bibinfo{person}{Paul Lukowicz}.} \bibinfo{year}{2008}\natexlab{}.
\newblock \showarticletitle{On body capacitive sensing for a simple touchless user interface}. In \bibinfo{booktitle}{\emph{2008 5th International Summer School and Symposium on Medical Devices and Biosensors}}. IEEE, \bibinfo{pages}{113--116}.
\newblock


\bibitem[Choi and Kim(2017)]%
        {choi2017driver}
\bibfield{author}{\bibinfo{person}{Minho Choi} {and} \bibinfo{person}{Sang~Woo Kim}.} \bibinfo{year}{2017}\natexlab{}.
\newblock \showarticletitle{Driver's movement monitoring system using capacitive ECG sensors}. In \bibinfo{booktitle}{\emph{2017 IEEE 6th Global Conference on Consumer Electronics (GCCE)}}. IEEE, \bibinfo{pages}{1--2}.
\newblock


\bibitem[Cohn et~al\mbox{.}(2012)]%
        {cohn2012ultra}
\bibfield{author}{\bibinfo{person}{Gabe Cohn}, \bibinfo{person}{Sidhant Gupta}, \bibinfo{person}{Tien-Jui Lee}, \bibinfo{person}{Dan Morris}, \bibinfo{person}{Joshua~R Smith}, \bibinfo{person}{Matthew~S Reynolds}, \bibinfo{person}{Desney~S Tan}, {and} \bibinfo{person}{Shwetak~N Patel}.} \bibinfo{year}{2012}\natexlab{}.
\newblock \showarticletitle{An ultra-low-power human body motion sensor using static electric field sensing}. In \bibinfo{booktitle}{\emph{Proceedings of the 2012 ACM conference on ubiquitous computing}}. \bibinfo{pages}{99--102}.
\newblock


\bibitem[Erickson et~al\mbox{.}(2018)]%
        {erickson2018tracking}
\bibfield{author}{\bibinfo{person}{Zackory Erickson}, \bibinfo{person}{Maggie Collier}, \bibinfo{person}{Ariel Kapusta}, {and} \bibinfo{person}{Charles~C Kemp}.} \bibinfo{year}{2018}\natexlab{}.
\newblock \showarticletitle{Tracking human pose during robot-assisted dressing using single-axis capacitive proximity sensing}.
\newblock \bibinfo{journal}{\emph{IEEE Robotics and Automation Letters}} \bibinfo{volume}{3}, \bibinfo{number}{3} (\bibinfo{year}{2018}), \bibinfo{pages}{2245--2252}.
\newblock


\bibitem[Forster(1974)]%
        {forster1974measurement}
\bibfield{author}{\bibinfo{person}{IC Forster}.} \bibinfo{year}{1974}\natexlab{}.
\newblock \showarticletitle{Measurement of patient body capacitance and a method of patient isolation in mains environments}.
\newblock \bibinfo{journal}{\emph{Medical and biological engineering}} \bibinfo{volume}{12}, \bibinfo{number}{5} (\bibinfo{year}{1974}), \bibinfo{pages}{730--732}.
\newblock


\bibitem[Grosse-Puppendahl et~al\mbox{.}(2012)]%
        {grosse2012enhancing}
\bibfield{author}{\bibinfo{person}{Tobias Grosse-Puppendahl}, \bibinfo{person}{Eugen Berlin}, {and} \bibinfo{person}{Marko Borazio}.} \bibinfo{year}{2012}\natexlab{}.
\newblock \showarticletitle{Enhancing accelerometer-based activity recognition with capacitive proximity sensing}. In \bibinfo{booktitle}{\emph{International Joint Conference on Ambient Intelligence}}. Springer, \bibinfo{pages}{17--32}.
\newblock


\bibitem[Grosse-Puppendahl et~al\mbox{.}(2017)]%
        {grosse2017finding}
\bibfield{author}{\bibinfo{person}{Tobias Grosse-Puppendahl}, \bibinfo{person}{Christian Holz}, \bibinfo{person}{Gabe Cohn}, \bibinfo{person}{Raphael Wimmer}, \bibinfo{person}{Oskar Bechtold}, \bibinfo{person}{Steve Hodges}, \bibinfo{person}{Matthew~S Reynolds}, {and} \bibinfo{person}{Joshua~R Smith}.} \bibinfo{year}{2017}\natexlab{}.
\newblock \showarticletitle{Finding common ground: A survey of capacitive sensing in human-computer interaction}. In \bibinfo{booktitle}{\emph{Proceedings of the 2017 CHI conference on human factors in computing systems}}. \bibinfo{pages}{3293--3315}.
\newblock


\bibitem[Heilbron(1966)]%
        {heilbron1966gm}
\bibfield{author}{\bibinfo{person}{John~L Heilbron}.} \bibinfo{year}{1966}\natexlab{}.
\newblock \showarticletitle{GM Bose: The Prime Mover in the Invention of the Leyden Jar?}
\newblock \bibinfo{journal}{\emph{Isis}} \bibinfo{volume}{57}, \bibinfo{number}{2} (\bibinfo{year}{1966}), \bibinfo{pages}{264--267}.
\newblock


\bibitem[Ianov et~al\mbox{.}(2012)]%
        {ianov2012development}
\bibfield{author}{\bibinfo{person}{Alexsandr~Igorevitch Ianov}, \bibinfo{person}{Hiroaki Kawamoto}, {and} \bibinfo{person}{Yoshiyuki Sankai}.} \bibinfo{year}{2012}\natexlab{}.
\newblock \showarticletitle{Development of a capacitive coupling electrode for bioelectrical signal measurements and assistive device use}. In \bibinfo{booktitle}{\emph{2012 ICME International Conference on Complex Medical Engineering (CME)}}. IEEE, \bibinfo{pages}{593--598}.
\newblock


\bibitem[Jonassen(1998)]%
        {jonassen1998human}
\bibfield{author}{\bibinfo{person}{Niels Jonassen}.} \bibinfo{year}{1998}\natexlab{}.
\newblock \showarticletitle{Human body capacitance: static or dynamic concept?[ESD]}. In \bibinfo{booktitle}{\emph{Electrical Overstress/Electrostatic Discharge Symposium Proceedings. 1998 (Cat. No. 98TH8347)}}. IEEE, \bibinfo{pages}{111--117}.
\newblock


\bibitem[Kang et~al\mbox{.}(2024)]%
        {kang2024device}
\bibfield{author}{\bibinfo{person}{Pixi Kang}, \bibinfo{person}{Julian Moosmann}, \bibinfo{person}{Sizhen Bian}, {and} \bibinfo{person}{Michele Magno}.} \bibinfo{year}{2024}\natexlab{}.
\newblock \showarticletitle{On-Device Training Empowered Transfer Learning For Human Activity Recognition}.
\newblock \bibinfo{journal}{\emph{arXiv preprint arXiv:2407.03644}} (\bibinfo{year}{2024}).
\newblock


\bibitem[Kuik(2004)]%
        {kuik2004digital}
\bibfield{author}{\bibinfo{person}{Sel-Vin Kuik}.} \bibinfo{year}{2004}\natexlab{}.
\newblock \bibinfo{title}{A Digital Theremin}.
\newblock
\newblock


\bibitem[Lissmann and Machin(1958)]%
        {lissmann1958mechanism}
\bibfield{author}{\bibinfo{person}{Hans~W Lissmann} {and} \bibinfo{person}{Ken~E Machin}.} \bibinfo{year}{1958}\natexlab{}.
\newblock \showarticletitle{The mechanism of object location in Gymnarchus niloticus and similar fish}.
\newblock \bibinfo{journal}{\emph{Journal of Experimental Biology}} \bibinfo{volume}{35}, \bibinfo{number}{2} (\bibinfo{year}{1958}), \bibinfo{pages}{451--486}.
\newblock


\bibitem[Liu et~al\mbox{.}(2022)]%
        {liu2022non}
\bibfield{author}{\bibinfo{person}{Mengxi Liu}, \bibinfo{person}{Sizhen Bian}, {and} \bibinfo{person}{Paul Lukowicz}.} \bibinfo{year}{2022}\natexlab{}.
\newblock \showarticletitle{Non-contact, real-time eye blink detection with capacitive sensing}. In \bibinfo{booktitle}{\emph{Proceedings of the 2022 ACM International Symposium on Wearable Computers}}. \bibinfo{pages}{49--53}.
\newblock


\bibitem[Matthies et~al\mbox{.}(2017)]%
        {matthies2017earfieldsensing}
\bibfield{author}{\bibinfo{person}{Denys~JC Matthies}, \bibinfo{person}{Bernhard~A Strecker}, {and} \bibinfo{person}{Bodo Urban}.} \bibinfo{year}{2017}\natexlab{}.
\newblock \showarticletitle{Earfieldsensing: A novel in-ear electric field sensing to enrich wearable gesture input through facial expressions}. In \bibinfo{booktitle}{\emph{Proceedings of the 2017 CHI Conference on Human Factors in Computing Systems}}. \bibinfo{pages}{1911--1922}.
\newblock


\bibitem[Nikitin(2012)]%
        {nikitin2012leon}
\bibfield{author}{\bibinfo{person}{Pavel Nikitin}.} \bibinfo{year}{2012}\natexlab{}.
\newblock \showarticletitle{Leon theremin (lev termen)}.
\newblock \bibinfo{journal}{\emph{IEEE Antennas and Propagation Magazine}} \bibinfo{volume}{54}, \bibinfo{number}{5} (\bibinfo{year}{2012}), \bibinfo{pages}{252--257}.
\newblock


\bibitem[Osoinach(2007)]%
        {osoinach2007proximity}
\bibfield{author}{\bibinfo{person}{Bryce Osoinach}.} \bibinfo{year}{2007}\natexlab{}.
\newblock \showarticletitle{Proximity capacitive sensor technology for touch sensing applications}.
\newblock \bibinfo{journal}{\emph{Freescale White Paper}}  \bibinfo{volume}{12} (\bibinfo{year}{2007}).
\newblock


\bibitem[Schilk et~al\mbox{.}(2022)]%
        {VitalPod}
\bibfield{author}{\bibinfo{person}{Philipp Schilk}, \bibinfo{person}{Kanika Dheman}, {and} \bibinfo{person}{Michele Magno}.} \bibinfo{year}{2022}\natexlab{}.
\newblock \showarticletitle{VitalPod: A Low Power In-Ear Vital Parameter Monitoring System}. In \bibinfo{booktitle}{\emph{2022 18th International Conference on Wireless and Mobile Computing, Networking and Communications (WiMob)}}. \bibinfo{pages}{94--99}.
\newblock
\urldef\tempurl%
\url{https://doi.org/10.1109/WiMob55322.2022.9941646}
\showDOI{\tempurl}


\bibitem[Shinagawa et~al\mbox{.}(2004)]%
        {shinagawa2004near}
\bibfield{author}{\bibinfo{person}{Mitsuru Shinagawa}, \bibinfo{person}{Masaaki Fukumoto}, \bibinfo{person}{Katsuyuki Ochiai}, {and} \bibinfo{person}{Hakaru Kyuragi}.} \bibinfo{year}{2004}\natexlab{}.
\newblock \showarticletitle{A near-field-sensing transceiver for intrabody communication based on the electrooptic effect}.
\newblock \bibinfo{journal}{\emph{IEEE Transactions on instrumentation and measurement}} \bibinfo{volume}{53}, \bibinfo{number}{6} (\bibinfo{year}{2004}), \bibinfo{pages}{1533--1538}.
\newblock


\bibitem[Skeldon et~al\mbox{.}(1998)]%
        {skeldon1998physics}
\bibfield{author}{\bibinfo{person}{Kenneth~D Skeldon}, \bibinfo{person}{Lindsay~M Reid}, \bibinfo{person}{Viviene McInally}, \bibinfo{person}{Brendan Dougan}, {and} \bibinfo{person}{Craig Fulton}.} \bibinfo{year}{1998}\natexlab{}.
\newblock \showarticletitle{Physics of the Theremin}.
\newblock \bibinfo{journal}{\emph{American Journal of Physics}} \bibinfo{volume}{66}, \bibinfo{number}{11} (\bibinfo{year}{1998}), \bibinfo{pages}{945--955}.
\newblock


\bibitem[Smith(1996)]%
        {smith1996field}
\bibfield{author}{\bibinfo{person}{Joshua~R. Smith}.} \bibinfo{year}{1996}\natexlab{}.
\newblock \showarticletitle{Field mice: Extracting hand geometry from electric field measurements}.
\newblock \bibinfo{journal}{\emph{IBM systems journal}} \bibinfo{volume}{35}, \bibinfo{number}{3.4} (\bibinfo{year}{1996}), \bibinfo{pages}{587--608}.
\newblock


\bibitem[Suh et~al\mbox{.}(2023)]%
        {suh2023worker}
\bibfield{author}{\bibinfo{person}{Sungho Suh}, \bibinfo{person}{Vitor~Fortes Rey}, \bibinfo{person}{Sizhen Bian}, \bibinfo{person}{Yu-Chi Huang}, \bibinfo{person}{Jo{\v{z}}e~M Ro{\v{z}}anec}, \bibinfo{person}{Hooman~Tavakoli Ghinani}, \bibinfo{person}{Bo Zhou}, {and} \bibinfo{person}{Paul Lukowicz}.} \bibinfo{year}{2023}\natexlab{}.
\newblock \showarticletitle{Worker activity recognition in manufacturing line using near-body electric field}.
\newblock \bibinfo{journal}{\emph{IEEE Internet of Things Journal}} \bibinfo{volume}{11}, \bibinfo{number}{7} (\bibinfo{year}{2023}), \bibinfo{pages}{11554--11565}.
\newblock


\bibitem[Takiguchi et~al\mbox{.}(2007)]%
        {takiguchi2007human}
\bibfield{author}{\bibinfo{person}{Kiyoaki Takiguchi}, \bibinfo{person}{Takayuki Wada}, {and} \bibinfo{person}{Shigeki Toyama}.} \bibinfo{year}{2007}\natexlab{}.
\newblock \showarticletitle{Human body detection that uses electric field by walking}.
\newblock \bibinfo{journal}{\emph{Journal of Advanced Mechanical Design, Systems, and Manufacturing}} \bibinfo{volume}{1}, \bibinfo{number}{3} (\bibinfo{year}{2007}), \bibinfo{pages}{294--305}.
\newblock


\bibitem[Tang and Mandal(2019)]%
        {tang2019indoor}
\bibfield{author}{\bibinfo{person}{Xinyao Tang} {and} \bibinfo{person}{Soumyajit Mandal}.} \bibinfo{year}{2019}\natexlab{}.
\newblock \showarticletitle{Indoor occupancy awareness and localization using passive electric field sensing}.
\newblock \bibinfo{journal}{\emph{IEEE Transactions on Instrumentation and Measurement}} \bibinfo{volume}{68}, \bibinfo{number}{11} (\bibinfo{year}{2019}), \bibinfo{pages}{4535--4549}.
\newblock


\bibitem[Yama et~al\mbox{.}(2007)]%
        {yama2007development}
\bibfield{author}{\bibinfo{person}{Yoshihiro Yama}, \bibinfo{person}{Akinori Ueno}, {and} \bibinfo{person}{Yoshinori Uchikawa}.} \bibinfo{year}{2007}\natexlab{}.
\newblock \showarticletitle{Development of a wireless capacitive sensor for ambulatory ECG monitoring over clothes}. In \bibinfo{booktitle}{\emph{2007 29th Annual International Conference of the IEEE Engineering in Medicine and Biology Society}}. IEEE, \bibinfo{pages}{5727--5730}.
\newblock


\bibitem[Zhao(2018)]%
        {zhao2018review}
\bibfield{author}{\bibinfo{person}{Jingdong Zhao}.} \bibinfo{year}{2018}\natexlab{}.
\newblock \showarticletitle{A review of wearable IMU (inertial-measurement-unit)-based pose estimation and drift reduction technologies}. In \bibinfo{booktitle}{\emph{Journal of Physics: Conference Series}}, Vol.~\bibinfo{volume}{1087}. IOP Publishing, \bibinfo{pages}{042003}.
\newblock


\end{thebibliography}


\appendix




\end{document}